\documentclass[preprint,prb,showpacs,superscriptaddress,
draft,amsmath,amssymb]{revtex4}

\usepackage{graphicx}
\usepackage{dcolumn}
\usepackage{bm}

\makeatletter

\def\lsim{\compoundrel<\over\sim}
\def\compoundrel#1\over#2{\mathpalette\compoundreL{{#1}\over{#2}}}
\def\compoundreL#1#2{\compoundREL#1#2}
\def\compoundREL#1#2\over#3{\mathrel
    {\vcenter{\hbox{$\m@th\buildrel{#1#2}\over{#1#3}$}}}}
\makeatother

\begin{document}

\title{Doping-dependence of nodal quasiparticle properties in
high-$T_{\rm c}$ cuprates studied by laser-excited angle-resolved 
photoemission spectroscopy}

\author{K. \surname{Ishizaka}}
\affiliation{Institute for Solid State Physics, University of Tokyo,
 Kashiwa, Chiba 277-8581, Japan}

\author{T. \surname{Kiss}}
\affiliation{Institute for Solid State Physics, University of Tokyo,
 Kashiwa, Chiba 277-8581, Japan}

\author{S. \surname{Izumi}}
\affiliation{Institute for Solid State Physics, University of Tokyo,
 Kashiwa, Chiba 277-8581, Japan}

\author{M. \surname{Okawa}}
\affiliation{Institute for Solid State Physics, University of Tokyo,
 Kashiwa, Chiba 277-8581, Japan}

\author{T. \surname{Shimojima}}
\affiliation{Institute for Solid State Physics, University of Tokyo,
 Kashiwa, Chiba 277-8581, Japan}

\author{A. Chainani}
\affiliation{The Institute of Physical and Chemical Research (RIKEN),
Sayo-gun, Hyogo 679-5143, Japan}

\author{T. Togashi}
\affiliation{The Institute of Physical and Chemical Research (RIKEN),
Sayo-gun, Hyogo 679-5143, Japan}

\author{S. Watanabe}
\affiliation{Institute for Solid State Physics, University of Tokyo,
 Kashiwa, Chiba 277-8581, Japan}

\author{C.-T. Chen}
\affiliation{Beijing Center for Crystal R\&D, Chinese Academy of Science,
Zhongguancun, Beijing 100080, China}

\author{X. Y. \surname{Wang}}
\affiliation{Beijing Center for Crystal R\&D, Chinese Academy of Science,
Zhongguancun, Beijing 100080, China}

\author{T. Mochiku}
\affiliation{National Institute for Material Science, 
Tsukuba, Ibaraki 305-0047, Japan}

\author{T. Nakane}
\affiliation{National Institute for Material Science, 
Tsukuba, Ibaraki 305-0047, Japan}

\author{K. Hirata}
\affiliation{National Institute for Material Science, 
Tsukuba, Ibaraki 305-0047, Japan}

\author{S. \surname{Shin}}
\affiliation{Institute for Solid State Physics, University of Tokyo,
 Kashiwa, Chiba 277-8581, Japan}
\affiliation{The Institute of Physical and Chemical Research (RIKEN),
Sayo-gun, Hyogo 679-5143, Japan}

\begin{abstract}

We investigate the doping dependent low energy, low temperature ($T$ = 5 K) 
properties of nodal quasiparticles 
in the $d$-wave superconductor 
Bi$_{2.1}$Sr$_{1.9}$CaCu$_2$O$_{8+\delta}$ (Bi2212).
By utilizing ultrahigh resolution
laser-excited angle-resolved photoemission spectroscopy, 
we obtain precise band dispersions near $E_{ F}$, mean free paths  
and scattering rates ($\Gamma$) of quasiparticles. 
For optimally and overdoped, 
we obtain very sharp quasiparticle peaks of 8 meV and
6 meV full-width at half-maximum, respectively, in accord with 
terahertz conductivity.
For all doping levels, we find the energy-dependence of 
$\Gamma \sim |\omega |$, while  $\Gamma$($\omega =0$) 
shows 
a monotonic increase from overdoping to underdoping. 
The doping dependence suggests the role of electronic 
inhomogeneity on 
the nodal quasiparticle scattering at low temperature 
(5 K $\lsim 0.07T_{\rm c}$), pronounced in the underdoped region.

\end{abstract}

\maketitle



Extensive studies over the past two decades have indicated a $d_{x^2-y^2}$
symmetry of the 
superconducting gap in the Bi-based high-$T_{\rm c}$ cuprates. \cite{tsuei00,damascelli03} 
The $d_{x^2-y^2}$-wave superconducting state is characterized by
line nodes along the ($\pm \pi/2, \pm \pi/2$) direction, 
and allow the existence of quasiparticles crossing the Fermi level even in the 
gapped superconducting ground state. 
Early in the literature, theoretical studies have discussed the behavior of nodal quasiparticles in terms of 
massless Fermions with a Dirac-cone dispersion.  
 \cite{nersesyan94}
By investigating such nodal quasiparticles which are responsible for the low energy 
excitations, one can acquire insight into the characteristic features of the scattering 
and interaction mechanisms remaining in the superconducting state.
Experimentally, the properties of nodal quasiparticles 
in the superconducting state can be 
obtained from transport measurements, 
such as thermal Hall conductivity \cite{zhang01},  
microwave and terahertz conductivity. \cite{hosseini99,corson00}  
These studies in YBa$_2$Cu$_3$O$_{7}$ (Y123) had 
revealed a rapid increase of the quasiparticle 
lifetime on cooling across the superconducting transition, 
\cite{hosseini99,zhang01} 
and the mean-free-path obtained by the thermal Hall conductivity 
at the lowest temperature attains a value of $\sim $1 $\mu $m. \cite{zhang01}
On the other hand, terahertz conductivity in Bi2212 shows a 
scattering rate with rather high values and a slower 
$T$-linear dependence below $T_{\rm c}$, compared to Y123. \cite{corson00}
Such differences have been discussed in terms of possible 
impurity effects, which may accompany 
phase fluctuations, charge and superfluid density inhomogeneity, etc.

Angle-resolved photoemission spectroscopy (ARPES)
is a powerful method to investigate the momentum-resolved nodal quasiparticle 
properties, both, as a function of temperature and energy ($\omega$).
The energy position of the quasiparticle peak in ARPES 
represents the band dispersion, 
whereas the width indicates the scattering rate $\Gamma$. 
Recent studies have carefully probed the ``kink" structure at around 
60 meV in the quasiparticle dispersion, now commonly observed in 
hole-doped cuprates. \cite{lanzara02,damascelli03} 
The kink accompanies a rapid decrease of the scattering rate, 
and is discussed as the renormalization of a bosonic mode, derived from 
phonons or magnons. \cite{gweon04,douglas07,terashima06} 
There are important studies \cite{damascelli03,kordyuk04,valla99} 
that discuss the nature of the scattering mechanism 
of the electrons from the $\omega $ dependence 
of the scattering rate, {\it e.g.} is it a 
Fermi liquid ? ( with $\Gamma \propto \omega ^2$), or 
a marginal Fermi liquid ? (with $\Gamma \propto |\omega |$) \cite{varma89}. 
Due to limitations of momentum- and energy-resolutions, 
however, the low temperature $\Gamma $ at $\omega =0$ (i.e. the Fermi level) 
has not been  
obtained accurately enough to discuss the
nodal quasiparticle properties in terms of unique elementary 
excitations in the $d$-wave superconducting state. 
In particular, to date, the discrepancies obtained in its values by 
transport measurements and ARPES remains to be conclusively clarified. 
Accordingly, its doping dependence at low energy and low temperature 
is also not yet 
clarified and remains an open question.
In this study, we investigate the character of nodal quasiparticles of Bi2212 as a function of doping, 
by VUV laser-excited ARPES measurements. 
By using a low-energy ($h \nu = 6.994$ eV) and 
coherent light source, we can achieve high energy- and momentum- 
resolutions ($\lsim 1$ meV and 0.0014 \AA), 
as well as bulk sensitivity ($\sim 100$ \AA).
From our measurements, 
we obtained very sharp quasiparticle peak of 
6 meV full-width at half-maximum for an
overdoped sample.
The scattering rate obtained from the 
quasiparticle peak width shows a clear doping dependence, 
increasing monotonically from the overdoped to the underdoped region,
reminiscent of the pseudogap phenomenon. 
It possibly arises from the electronic inhomogeneity effect, 
as is known from scanning tunneling microscopy (STM) 
measurements.

ARPES measurements were performed using a system constructed with 
a VG-Scienta R4000 electron analyzer and an ultra-violet ($h\nu =6.994$ eV) 
laser for the incident light. \cite{kiss05} 
The temperature was precisely controlled
down to 5 K using a flow-type He liquid refrigerator. 
The pressure of the chamber was below $\sim 5\times 10^{-11}$ Torr
throughout all the measurements.
The energy and angular resolutions were
$E_{\rm res} = $1.0 meV and $\theta_{\rm res} = 0.1^{\circ} $.
The Fermi level ($E_{ F}$) of the sample was referred to that of 
a Au film 
evaporated on the sample substrate, with an accuracy of $\pm $ 0.1 meV.
High quality single crystals of Bi2212 were grown by 
traveling-solvent floating-zone method. \cite{mochiku97}
$T_{\rm c}$ of the samples from underdoped to overdoped region 
were 72 K (UN72K), 80 K (UN80K), 90 K (OP90K), 85 K (OV85K), 
and 73 K (OV73K).
The samples were cleaved {\it in situ} to obtain clean surfaces.
All the data presented in this work were obtained at 5 K.


Figure \ref{fig1}(a) shows 
the ARPES intensity image along the 
nodal direction [$\Gamma $ - Y: see Fig. \ref{fig1}(b) for a 
schematic Brillouin zone] from OP90K. 
As is well-known, \cite{damascelli03} 
the ``kink" 
structure reflecting the band renormalization at $\sim$60 meV is clearly observed 
in the raw data. 
To discuss the detailed features appearing 
in the band dispersion, we performed a 
fitting analysis of momentum distribution curves (MDC: 
momentum-profile of the ARPES intensity 
at a fixed binding energy) using the sum of a Lorenzian peak 
and a constant background, as a function of binding energy.
The obtained peak positions are overlaid as a red curve 
in Fig. \ref{fig1}(a), 
and the full-width-half-maximum (FWHM) 
$\Delta k$ is plotted as a function of binding energy ($E_{B}$) 
in Fig. \ref{fig1}(c), respectively.
$\Delta k$ thus obtained can be  approximately described by  
$\Delta k \approx 2{\rm Im}\Sigma /v_0$, where 
$v_0$ and $\Sigma$ correspond to the 
bare Fermi velocity and the self-energy of 
electrons.\cite{damascelli03}
The 60 meV ``kink" is easily identified as the knee in the 
band dispersion.  
In addition, a rapid decrease of
$\Delta k$ at lower binding energies is observed, which has been extensively 
discussed as a 
renormalization effect due to a bosonic mode. 
While these ``kink" properties are very similar to those 
reported previously by higher-energy ARPES results, 
the absolute value of the linewidth is obviously sharper here ;
$\Delta k \sim  0.01$ \AA $^{-1}$.
Similar results were recently reported by ARPES 
experiments using a laser source \cite{koralek06} as well as low-energy synchrotron light, \cite{yamasaki06} 
demonstrating the effectiveness and impact of low-energy ARPES for 
high resolution measurements, not only in energy but also in 
momentum.

In Fig. \ref{fig2}(a-c), the results of the ultrahigh 
energy- and momentum- 
resolution measurement are shown for UN72K, OP90K, and OV73K. 
A clear decrease of the linewidth on increasing the doping level is 
readily visible from the ARPES images. 
The energy distribution curves 
(EDCs: energy-spectrum of the ARPES intensity 
at a fixed momentum) from OV73K, shown in Fig. \ref{fig2}(d), indicate  
the continuous sharpening of the quasiparticle peak even at binding energies below $E_{\rm B}$ = 10 meV.
The quasiparticle peak eventually becomes a sharp  
peak with $\sim $ 6 meV FWHM at $k=k_{ F}$ [red curve in Fig. \ref{fig2}(d)]. 
By fitting the MDCs by Lorenzians, we obtained clear band dispersions 
which are overlaid as red curves on the images in Fig. \ref{fig2}(a-c). 
The FWHM $\Delta k$ as a function of $E_{\rm B}$ is plotted in 
Fig. \ref{fig2}(e) for UN72K, OP90K, and OV73K. 
Although the magnitudes of the FWHM $\Delta k$ are quite similar for 
OP90K and OV73K (with small differences only at
energies below 20 meV), they are significantly higher for the UN72K sample. 
In spite of these differences,
an overall quasilinear $\omega $ dependence at low energies is observed for 
all the samples.
Another important observation is that $\Delta k$ at $E_{ F}$ is  
significantly 
doping dependent; $\Delta k (0)$ = 0.014 \AA $^{-1}$ for UN72K, reduces 
to 0.0062 \AA $^{-1}$ for OP90K, 
and reduces further to 
0.0039 \AA $^{-1}$ for OV73K. 
The corresponding MDCs at $E_{ F}$ and the fitted Lorenzian curves 
are shown in Fig. \ref{fig3} (c). We can estimate the mean free 
path $l_{\rm mfp}$ of the nodal 
quasiparticles from 
$l_{\rm mfp} = \Delta k (0) ^{-1} $. The estimated values are 
$l_{\rm mfp} > 260$ \AA ~(OV73K), 
160 \AA ~(OP90K), and 70 \AA ~(UN72K), respectively.
The obtained values of $\Delta k$ and $l_{\rm mfp}$ for all the samples 
are shown in Table \ref{table1}.
It is noted that the results for the OP90K sample are in very good 
agreement with 
the work of Koralek {\it et al}. \cite{koralek06} 
Here we would also like to point out that while recent studies have 
shown the existence of bilayer splitting at the 
nodal point,\cite{kordyuk04PRB,yamasaki06,valla06} the MDC widths obtained for 
OP90K and OV73K in the present work are smaller than the reported 
bilayer splitting of 0.0075 \AA $^{-1}$. This indicates that, the
present results on OP90K and OV73K, and that of Koralek {\it et al}. 
for optimally doped Bi2212, 
measures only one of the bilayer split bands, presumably the antibonding 
band, at the photon energies of the incident laser used in the studies. 
The spectral weight measured for the bonding and antibonding band 
is known to strongly depend on the incident 
photon energy. \cite{kordyuk04PRB}

Now we compare the EDC at $k_{ F}$ 
among the different doping samples. 
The EDCs are shown in Fig. \ref{fig3}(a). 
While the EDC from OV73K shows a clear and sharp peak with very little 
background, the EDC peak from UN72K apparently tends to become broader. 
To get rid of the Fermi-Dirac distribution effect at $E_F$ and estimate the 
FWHM of the peak from EDCs, we symmetrized them 
at $E_{ F}$ and plotted them in Fig. \ref{fig3}(b). 
The symmetrized EDCs can be well fitted with a single Lorenzian function, 
as shown by the broken curves in Fig. \ref{fig3}(b).
The FWHM obtained from the symmetrized EDCs are 
12.5$\pm 0.6$ (UN72K), 8.4$\pm 0.8$ (OP90K) and 6.0$\pm 0.8$ meV (OV73K), 
respectively, where the errors mainly arise from 
the determination of $k_F$  (see Table \ref{table1} for all samples). 
The FWHM of an EDC peak corresponds to the scattering rate 
$\Gamma   =h/\tau  \approx v_0/v_{ F} {\rm Im} \Sigma $,
where  $\tau $ and $v_{ F}$ correspond to the lifetime and the 
renormalized Fermi velocity of quasiparticles.
The relation between EDC and MDC peak FWHM 
can be expressed as 
$\Gamma \approx v_{F} \Delta k  $, where $v_{ F}$ can be estimated from
the gradient of the band dispersion in the vicinity of $E_{ F}$. 
By linearly approximating the dispersion at $10 {\rm meV} \leq E_{\rm B} 
 \leq 0$ meV obtained from the MDC analysis 
(red curves in Fig. \ref{fig2}(a-c)), 
we can get $v_{ F}$ within errors of about $\pm 10$\% 
and thus $\Gamma $ from MDCs in Fig. \ref{fig3}(c) 
as 15$\pm 1.4$ (UN72K), 8.1$\pm 0.6$ (OP90K), and 5.1$\pm 0.4$ meV (OV73K), 
which are very similar to the values obtained from 
EDCs (see Table \ref{table1}). 
Thus, we succeeded in obtaining the coherent nodal 
quasiparticle component from 
ARPES and estimating its scattering rate $\Gamma$.
Here, it is important to compare our results with that from
previous transport measurements.
The quasiparticle scattering rate obtained from terahertz conductivity 
using a nearly optimally-doped sample ($T_{\rm c} =85$ K) \cite{corson00} 
is $1/\tau_{\rm QP} \sim 3$ THz $\sim$ 12 meV at the lowest temperature 
of measurement, 12 K. By assuming a simple linear  
extrapolation, the scattering rate should be 
$1/\tau_{\rm QP} \sim 2$ THz $\sim$ 8 meV at around 5 K. 
This value is equivalent with our result from 
OP90K (8.1 - 8.4 meV), indicating full 
consistency among two completely different probes. 
While not yet reported, a similar doping dependence of $1/\tau_{\rm QP}$
can also be expected in terahertz conductivity measurements on Bi2212.

The doping dependence of the FWHM $\Gamma$ thus obtained is plotted in 
Fig. \ref{fig4}, with that of $T_{\rm c}$. 
$\Gamma$ shows a monotonic increase from overdoping to underdoping.
Such tendency itself is very similar to that of the well-known 
pseudogap in ($\pi$,0) region, \cite{timusk99} 
being intensively discussed to date as the 
manifestation of antiferromagnetic spin fluctuation, 
charge order, phase fluctuation, and so on.
Very recently, systematic investigations of the
electronic structures by Raman scattering \cite{tacon06} and 
ARPES \cite{tanaka06,kondo07} measurements 
have revealed that there is a strong dichotomy among the 
near-nodal (Fermi arc) and the antinodal (pseudogap) regions. 
The Fermi arc region shows the opening of a 
well-defined $d$-wave  superconducting gap with coherent quasiparticles 
regardless of the doping level, 
while the pseudogap shows a monotonic increase on underdoping 
with the enhancement of the incoherent spectral shape. 
STM measurements also indicate the spatially inhomogeneous  
(pseudo)gap spectra arising in the underdoped samples, 
while keeping rather homogeneous coherent quasiparticle spectra for 
all doping levels. 
\cite{lang02,mcelroy05a,mcelroy03}
Our observation of the well-defined coherent peaks is basically in 
agreement with this picture. 
However, it also shows that the lifetime of the 
nodal quasiparticle at the center of the Fermi arc 
shows a significant doping dependence, when investigated with high energy and momentum resolution at $T$ = 5 K. 

The lifetime of the nodal quasiparticle can be affected by a 
number of scattering mechanisms.
In case of a metal with nearly flat density of states (DOS) near $E_{F}$, 
electron-electron scattering can 
provide the low-energy scattering rate ($\omega \to 0$) as 
Im$\Sigma \propto $ max($\omega ^2, T^2$) 
in a Fermi liquid, while the Im$\Sigma \propto $ max($|\omega|, T$) in a
marginal Fermi liquid picture. \cite{varma89} 
Similarly, the electron-phonon scattering gives 
Im$\Sigma \propto {\rm max}(\omega ^3, T^3)$.
In a $d$-wave superconducting state, they are completely 
modified reflecting the 
$|\omega|$-linear DOS near $E_F$. \cite{dahm05} 
The dissipation by electron-electron (or spin fluctuation) scattering is 
given by 
Im$\Sigma \propto {\rm max}(\omega ^3, T^3)$, and electron-phonon by
Im$\Sigma \propto {\rm max}(\omega ^4, T^4)$. 
In the low energy and temperature limit (appropriate for the present 
measurements 
performed at $T\lsim 0.07T_{\rm c}$ with $E_{\rm res} \sim 1$ meV), 
however, all the above scattering effects should become negligible.
In fact, the $\omega$ dependence of the Im$\Sigma \propto \Delta k$ shows 
quasilinear behavior, indicating that the above dissipation mechanisms 
are not dominant in this $T, \omega$ region. 
One of the remaining terms is the elastic impurity scattering mechanism, 
which can affect a $d$-wave superconducting state in a non-trivial way,
unlike in a normal metal. \cite{nersesyan94,lee93}
An in-plane impurity, for example, is known to  
strongly scatter the quasiparticles and 
create a localized state around the node appearing as an enhancement of 
Im$\Sigma (\omega)$ near $E_F$. \cite{lee93}
In case of the out-of-plane impurity which likely provides a weak 
forward scattering potential, on the other hand, 
Im$\Sigma $ of the quasiparticles can 
show a $|\omega|$-linear dependence at $\omega \to 0$ with 
the value of Im$\Sigma (\omega =0)$ depending on the scattering length of 
impurity potential and its concentration. \cite{dahm05} 
$\Delta k$ showing a $\omega $ dependence of convex upward structure at 
around $E_{\rm B}=10$ meV in OP90K may be due to such an
elastic scattering, as expected from calculations 
based on a BCS $d$-wave model. \cite{dahm05}
Recently, the interplay among the apical oxygen dopant, the 
gap inhomogeneity and quasiparticle interference has been 
observed in STM measurements. 
\cite{mcelroy05b,zhou07}
The weakening of the screening effect due to the reduction of 
carrier density 
may make the system more susceptible to impurity scattering, 
typically emerging as granular (typical size of 3 nm) superconductivity in 
underdoped samples \cite{lang02,mcelroy05a}. 
It is interesting that the mean free path of the nodal quasiparticle in 
UN72K is $l_{\rm mfp} = 70 $ \AA, a little bit longer than 
the reported size of the 
granular superconducting domains.
On the other hand, we cannot yet completely rule out the possibility of 
the quantum fluctuation enhancement in 
underdoped samples, such as phase fluctuation arising from the 
reduction of superfluid density, charge order fluctuation, 
{\it etc}, if they have very low energy scales ($\lsim $ 1 meV).
Future work including detailed temperature-dependent measurements 
are required to elucidate on these issues.

In conclusion, we have performed an ultrahigh resolution angle-resolved
photoemission spectroscopy measurement to elucidate the 
properties of nodal quasiparticles  in a $d$-wave superconductor.
The quasilinear $\omega $ dependence of the scattering rate $\Gamma $ 
obtained from MDCs suggest 
the role of an 
elastic scattering mechanism by out-of-plane impurities, such 
as oxygen dopants.
The value of $\Gamma $ at $\omega = 0$ for optimally and overdoped samples are 
8 meV and 6 meV, respectively, 
in a good correspondence with terahertz conductivity results.
The doping dependence of $\Gamma $ resembles that of the 
pseudogap, monotonically increasing up to 12 meV on approaching 
the underdoped region. 
The estimated mean free paths accordingly decreases from 260 {\AA} in 
an overdoped to 70 {\AA} in an underdoped sample.
This suggests role of electronic inhomogeneity on 
the nodal quasiparticle scattering at low temperature 
in underdoped samples, as is known from 
scanning tunneling microscopy (STM) 
measurements.

We thank Y. Yanase for valuable discussion.

\newpage

\begin{table}[htbp]
\begin{center}
\begin{tabular}{cc|cccc|c}\hline \hline
sample & ~~$ p$ ~~ & ~~$\Delta k $ { [\AA$^{-1}$]}~~ & ~~$l_{\rm mfp}$ { [\AA]}~~ &~~$v_F$ { [\rm eV\AA]}~~&~~$v_F \Delta k$ { [meV]}~~& ~~$\Gamma$ { [meV]}~~ \\ 
\hline 
UN72K & 0.11 & 0.0136 & 74  & 1.1 ($\pm 0.1$) & 15.0 ($\pm 1.4$) & 12.5 ($\pm 0.6$) \\
UN80K & 0.12 & 0.0136 & 74  & 1.1 ($\pm 0.1$) & 15.0 ($\pm 1.4$) & 11.7 ($\pm 0.6$) \\
OP90K & 0.16 & 0.0062 & 160 & 1.3 ($\pm 0.1$) &  8.1 ($\pm 0.6$) &  8.4 ($\pm 0.8$) \\ 
OV85K & 0.19 & 0.0075 & 130 & 1.0 ($\pm 0.1$) &  8.0 ($\pm 0.8$) &  7.2 ($\pm 0.6$) \\ 
OV73K & 0.21 & 0.0039 & 260 & 1.3 ($\pm 0.1$) &  5.1 ($\pm 0.4$) &  6.0 ($\pm 0.8$) \\ \hline \hline
\end{tabular}
\caption{Doping dependence of the nodal quasiparticle property at $\omega =0$.
$p$ is the hole concentration estimated from 
$T_{\rm c}=T_{\rm c}^{\rm max}  (1-82.6 (0.16-p)^2)$, 
$\Delta k$, $l_{\rm mfp}$, $v_F$, and $v_F \Delta k$ are 
FWHM, mean free path, Fermi velocity, and scattering rate 
obtained from MDC analysis, 
$\Gamma $ is the scattering rate obtained from 
EDC analysis. See the text for details. }
\label{table1}
\end{center}
\end{table}

\begin{figure}[htbp!]
\caption{
(color). (a) ARPES intensity image plot from Bi2212 OP90K. The 
red curve shows the band dispersion as obtained from the 
MDC peak positions.
(b) A schematic of the Brillouin zone of Bi2212. The 
red line shows the 
momentum-region of the measurement [along (0,0)-($\pi$,$\pi$)].
(c) $E_{\rm B}$-dependence of the MDC FWHM obtained from OP90K.
\label{fig1}
}
\end{figure}

\begin{figure}[htbp!]
\caption{
(color). (a)-(c)
Ultrahigh resolution ARPES intensity image plot from Bi2212 
UN72K, OP90K, and OV73K, respectively.
The red curves show the band dispersions obtained from 
MDC peak positions.
(d) EDC of taken from ARPES on OV73K (c).
(e) $E_{\rm B}$-dependence of
MDC FWHM obtained from UN72K (a), OP90K (b), and OV73K (c), 
respectively.
\label{fig2}
}
\end{figure}

\begin{figure}[htbp!]
\caption{
(color online). 
EDC at $k=k_{ F}$ (a) and that symmetrized at $E_{ F}$ (b) 
from ARPES on Bi2212 UN72K, OP90K, and OV73K [Fig. \ref{fig2}(a)-(c)]. 
(c) shows the MDC at $E=E_{ F}$ obtained from the 
same ARPES spectra. The curves in (b) and (c) are the 
fitting result by using a single Lorenzian function.
\label{fig3}
}
\end{figure}

\begin{figure}[htbp!]
\caption{
(color online). 
Doping dependence of the scattering rate $\Gamma$ estimated from 
the FWHM of EDC and MDC in Bi2212. 
The right axis shows the $T_{\rm c}$ of 
the measured samples with the relation to hole concentration $p$, 
$T_{\rm c}=T_{\rm c}^{\rm max}  (1-82.6 (0.16-p)^2)$.
\label{fig4}
}
\end{figure}


\begin{thebibliography}{99}


\bibitem{tsuei00} 
C. C. Tsuei and J. R. Kirtley,
Rev. Mod. Phys. {\bf 72}, 969 (2000).

\bibitem{damascelli03} 
A. Damascelli, Z. Hussain, Z.-X. Shen, 
Rev. Mod. Phys. {\bf 75}, 473 (2003).


\bibitem{nersesyan94} 
A. A. Nersesyan, A. M. Tsvelik, and F. Wenger, 
Phys. Rev. Lett. {\bf 72}, 2628 (1994).


\bibitem{zhang01} 
Y. Zhang, N. P. Ong, P. W. Anderson, D. A. Bonn, R. Liang, and 
W. N. Hardy, 
Phys. Rev. Lett. {\bf 86}, 890 (2001).

\bibitem{hosseini99} 
A. Hosseini, R. Harris, S. Kamal, P. Dosanjh, J. Preston, 
R. Liang, W. N. Hardy, and D. A. Bonn,
Phys. Rev. B {\bf 60}, 1349 (1999).


\bibitem{corson00} 
J. Corson, J. Orenstein, S. Oh, J. O'Donnell, and J. N. Eckstein,
Phys. Rev. Lett. {\bf 85}, 2569 (2000).




\bibitem{lanzara02} 
A. Lanzara, P. V. Nogdanov, X. J. Zhou, S. A. Kellar, D. L. Feng, 
E. D. Lu, T. Yoshida, H. Eisaki, A. Fujimori, K. Kishio, J.-I Shimoyama, 
T. Noda, S. Uchida, Z. Hussain, and Z.-X. Shen, 
Nature (London) {\bf 415}, 412 (2002).

\bibitem{gweon04} 
G.-H. Gweon, T. Sasagawa, S. Y. Zhou, J. Graf, H. Takagi, 
D.-H. Lee, and A. Lanzara, 
Nature (London) {\bf 430}, 187 (2004).


\bibitem{douglas07} 
J. F. Douglas, H. Iwasawa, Z. Sun, A. V. Fedrov, M. Ishikado, 
T. Saitoh, H. Eisaki, H. Bando, T. Iwase, A. Ino, M. Arita, K. Shimada, 
H. Namatame, M. Taniguchi, T. Masui, S. Tajima, K. Fujita, S. Uchida, 
Y. Aiura, and D. S. Dessau,  
Nature (London) {\bf 446}, 15 (2007).


\bibitem{terashima06} 
K. Terashima, H. Matsui, D. Hashimoto, T. Sato, T. Takahashi, H. Ding, 
T. Yamamoto, and K. Kadowaki, 
Nature Physics (London) {\bf 2}, 27 (2006).




\bibitem{varma89} 
C. M. Varma, P. B. Littlewood, S. Schmitt-Rink, E. Abrahams, and 
A. E. Ruckenstein,  
Phys. Rev. Lett. {\bf 63}, 1996 (1989).


\bibitem{kordyuk04} 
A. A. Kordyuk, S. V. Borisenko, A. Koitzsch, J. Fink, M. Knupfer, 
B. Buchner, H. Berger, G. Margaritondo, C. T. Lin, B. Keimer, S. Ono, 
and Y. Ando,
Phys. Rev. Lett. {\bf 92}, 257006 (2004).




\bibitem{valla99} 
T. Valla, A. V. Fedorov, P. D. Johnson, B. O. Wells, S. L. Hilbert, 
Q. Li, G. D. Gu, N. Koshizuka, 
Science {\bf 285}, 2110 (1999).


\bibitem{kiss05} 
T. Kiss, F. Kanetaka, T. Yokoya, T. Shimojima, K. Kanai, S. Shin, 
Y. Onuki, T. Togashi, C. Zhang, C. T. Chen, and S. Watanabe,
Phys. Rev. Lett. {\bf 94}, 057001 (2005).


\bibitem{mochiku97}
T. Mochiku, K. Hirata, and K. Kadowaki, 
Physica (Amsterdam) {\bf 282C-287C}, 475 (1997).






\bibitem{koralek06} 
J. D. Koralek, J. F. Douglas, N. C. Plumb, Z. Sun, A. V. Federov, 
M. M. Murnane, H. C. Kapteyn, S. T. Cundiff, Y. Aiura, K. Oka, 
H. Eisaki, and D. S. Dessau, 
Phys. Rev. Lett. {\bf 96}, 017005 (2006).


\bibitem{yamasaki06} 
T. Yamasaki, K. Yamazaki, A. Ino, M. Arita, H. Namatame, M. Taniguchi, 
A. Fujimori, Z.-X. Shen, M. Ishikado, and S. Uchida, 
cond-mat/0603006v1 (2006).

\bibitem{valla06}
T. Valla, T. E. Kidd, J. D. Rameau, H.-J. Noh, G. D. Gu, P. D. Johnson, 
H.-B. Yang, and H. Ding, 
Phys. Rev. B {\bf 73}, 184518 (2006).


\bibitem{kordyuk04PRB}
A. A. Kordyuk, S. V. Borisenko, A. N. Yaresko, S. -L. Drechsler, 
H. Rosner, T. K. Kim, A. Koitzsch, K. A. Nenkov, M. Knupfer, J. Fink, 
R. Follath, H. Berger, B. Keimer, S. Ono, and Y. Ando, 
Phys. Rev. B {\bf 70}, 214525 (2004).

\bibitem{timusk99}
T. Timusk and B. Statt,
Rep. Prog. Phys. {\bf 62}, 61 (1999).


\bibitem{opel00} 
M. Opel, R. Nemetschek, C. Hoffmann, R. Philipp, P. F. Muller, R. Hackl, 
I. Tutto, A. Erb, B. Revaz, E. Walker, H. Berger, and L. Forro,  
Phys. Rev. B {\bf 61}, 9752 (2000).


\bibitem{tacon06} 
M. L. Tacon, A. Sacuto, A. Georges, G. Kotliar, Y. Gallais, 
D. Colson, and A. Forget, 
Nature Physics (London) {\bf 2}, 537 (2006).



\bibitem{tanaka06} 
K. Tanaka, W. S. Lee, D. H. Lu, A. Fujimori, T. Hujii, 
Risdiana, I. Terasaki, D. J. Scalapino, T. P. Devereaux, 
Z. Hussain, and Z.-X. Shen,   
Science {\bf 314}, 1910 (2006).

\bibitem{kondo07}
T.Kondo, T.Takeuchi, A.Kaminski, S.Tsuda, and S.Shin,
Phys. Rev. Lett. {\bf 98}, 267004 (2007).


\bibitem{lang02} 
K. M. Lang, V. Madhavan, J. E. Hoffman, E. W. Hudson, H. Eisaki, 
S. Uchida, and J. C. Davis,  
Nature (London) {\bf 415}, 412 (2002).


\bibitem{mcelroy05a} 
K. McElroy, D.-H. Lee, J. E. Hoffman, K. M. Lang, J. Lee, E. W. Hudson, 
H. Eisaki, S. Uchida, and J. C. Davis,
Phys. Rev. Lett. {\bf 94}, 197005 (2005).


\bibitem{mcelroy03} 
K. McElroy, R. W. Simmonds, J. E. Hoffman, D.-H. Lee, J. Orenstein,
H. Eisaki, S. Uchida, and J. C. Davis,  
Nature (London) {\bf 422}, 592 (2003).



\bibitem{dahm05} 
T. Dahm, P. J. Hirschfeld, D. J. Scalapino, and L. Zhu, 
Phys. Rev. B {\bf 72}, 214512 (2005).

\bibitem{lee93} 
P. A. Lee,   
Phys. Rev. Lett. {\bf 71}, 1887 (1993).




\bibitem{mcelroy05b} 
K. McElroy, J. Lee, J. A. Slezak, D.-H. Lee, ,
H. Eisaki, S. Uchida, and J. C. Davis,  
Science {\bf 309}, 1048 (2005).

\bibitem{zhou07} 
S. Zhou, H. Ding, Z. Wang, 
Phys. Rev. Lett. {\bf 98}, 076401 (2007).






\end{thebibliography}
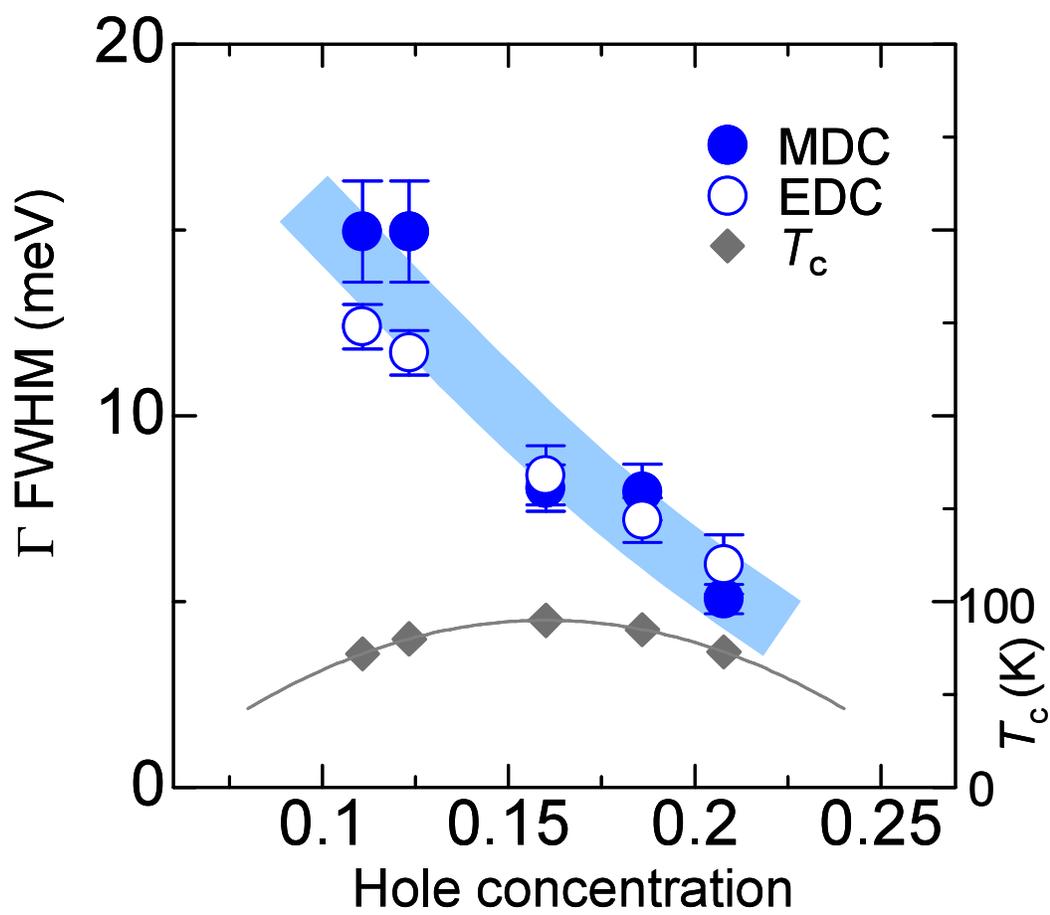
\end{document}